\documentclass[twocolumn,prl,showpacs]{revtex4-1}

\usepackage{graphicx,amssymb,amsmath}

\begin{document}

\title{Transition from wave turbulence to dynamical crumpling in vibrated elastic plates}
\author{Benjamin Miquel, Alexandros Alexakis}
\affiliation{Laboratoire de Physique Statistique, Ecole Normale Sup\'erieure, Universit\'e Pierre et Marie Curie, CNRS, 24 rue Lhomond, 75005 Paris, France.}
\author{Christophe Josserand}
\affiliation{Institut d'Alembert, UMR 7190, CNRS and UPMC (Paris 6), 4 place Jussieu, 75005 Paris, France}
\author{Nicolas Mordant}
\email[]{nicolas.mordant@ujf-grenoble.fr}
\affiliation{Laboratoire des Ecoulements G\'eophysiques et Industriels, Universit\'e Grenoble Alpes, Domaine Universitaire, BP53, 38041 Grenoble, France}
\affiliation{Institut Universitaire de France}


\begin{abstract}
We study the dynamical regime of wave turbulence of
a vibrated thin elastic plate based on experimental and
numerical observations. We focus our study to the strongly non linear regime described in a previous letter by N. Yokoyama \& M. Takaoka. At small forcing, a weakly non linear regime is compatible with the Weak Turbulence Theory when the dissipation is localized at high wavenumber. When the forcing intensity is increased, a strongly non linear regime emerges: singular structures dominate the dynamics at large scale whereas at small scales the weak turbulence is still present. A turbulence of singular structures, with folds and D-cones, develops that alters significantly the energy spectra and causes the emergence of intermittency.
\end{abstract}

\maketitle

A large ensemble of weakly non linear waves can develop a state of turbulence for which analytical derivations of the energy transfer can be performed:  the Weak (or Wave) Turbulence Theory (WTT). Because of the hypothesis of weak nonlinearities, the statistics remain close to Gaussian and the WTT enables to 
compute the so-called kinetic equation for the evolution of the spectral content of energy. The theory is appealing in the perspective of the theoretical study of turbulence in general because exact solutions of out of equilibrium dynamics can be derived analytically. These solutions show an energy flux between the large forcing scales and the small dissipation scales analogous to the Kolmogorov similarity solution for hydrodynamic turbulence~\cite{Zakharov,Newell,Nazarenko}.

Vibrating elastic plate are a fruitful model for wave turbulence study. WTT has been applied to this case in~\cite{During} and advanced measurements have been implemented that provided unprecedented results~\cite{Boudaoud, Mordant,Cobelli,epjb,Miquel,Miquel3}.
The normal deformation of a thin elastic plates follows the F\"oppl-Von Karman equations~\cite{Landau,Amabili}:
\begin{eqnarray}
\label{eq.1}
\rho \frac{\partial^2\zeta}{\partial t^2}&=&-\frac{Eh^2}{12(1-\nu^2)}\Delta^2\zeta+{\mathcal L}(\chi,\zeta),\\
\Delta^2\chi&=&-\frac{E}{2}{\mathcal L}(\zeta,\zeta),
\end{eqnarray}
where $\zeta$ is the transverse displacement, $\chi$ the Airy stress function, $h$ the thickness, $\rho$ the density, $E$ Young's modulus and $\nu$ the Poisson ratio. The operator ${\mathcal L}$ is bilinear symmetric, defined in cartesian coordinates by: ${\mathcal L}(f,g)=\frac{\partial^2 f}{\partial x^2}\frac{\partial^2 g}{\partial y^2}+\frac{\partial^2 f}{\partial y^2}\frac{\partial^2 g}{\partial x^2}-2\frac{\partial^2 f}{\partial x\partial y}\frac{\partial^2 g}{\partial x\partial y}$. 
The theory predicts an energy cascade from the forcing large scale to the small dissipative ones, the so-called Kolmogorov-Zakharov (KZ) spectrum, following for the energy density:
\begin{equation}
E_\zeta(k)=k \langle|\zeta_{\mathbf k}|^2\rangle=C  \frac{ h P^{1/3 } \rho^{2/3}}{(12(1-\nu^2))^{2/3}}   \frac{\ln^{1/3}(k_*/k)}{k^3},
\label{denspec}
\end{equation}
$\zeta_{\mathbf k}$ being the Fourier transform of the displacement, $P$ the energy flux within the cascade and $k^*$ a cut-off scale related to the dissipation
process~\cite{During}. 

While such scaling has been observed in numerical simulations where the dissipation is concentrated at small scales~\cite{During}, 
experiments have shown a different behavior following approximately $E_k \propto P^{0.6}k^{-3.5}$~\cite{Boudaoud,Mordant}. Such discrepancy between the WTT and the experiments is mostly due to the {\it real} dissipation that is present at every scale so that
no truly transparent window is present~\cite{Humbert,Miquel2}.\\

Recently, numerical studies have shown that a new strongly nonlinear regime emerges for intense forcing~\cite{Yokoyama}. In fact, when the forcing increases, one expects that the assumption of {\it weak} nonlinearities fails, either at large or small scales, leading {\it a priori} to a new dynamics~\cite{ZakNew,NNB}. Such 
regime is hard to handle and no general theory exists yet.
For instance, in systems allowing an inverse cascade, a condensate can develop that strongly alters the direct energy transfer~\cite{DuPiRi}. For water waves, new regimes of turbulence have been identified for varying forcing
amplitude~\cite{CoPr11}. 
Here we report experimental and numerical studies of this new regime. We show that intermittency appears at strong forcing and that it can be attributed to the existence of singular structures in the wave field.

\paragraph{Experiment}
The experimental setup and the presented dataset are the same as those of~\cite{Miquel3}. A stainless steel plate $2\times1$~m$^2$ and $h=0.4$~mm thick is held freely hanging from a beam. Vibrations are excited at 30~Hz by a electromagnetic shaker. 
The forcing intensity is tuned by changing the amplitude of the excitation. 
The deformation of the plate is measured using a high speed Fourier transform profilometry technique~\cite{Cobelli,Cobelli1} providing movies of the deformation over about 1~m$^2$ (i.e. half the total surface of the plate) that are resolved both in time and space. The recording frame rate is varied between 5000 and 10000 frames/s depending of the forcing intensity. 

\paragraph{Numerical simulation}
The F\"oppl-Von Karman equations (\ref{eq.1}) are simulated with forcing and dissipation by a pseudo spectral algorithm with second order Runge-
Kutta scheme and anti-aliasing~\cite{Miquel2} similar to that used in~\cite{During}. Resolutions are up to $768^2$ grid points to 
ensure a full development of the wave spectrum. Physical parameters are chosen to be comparable to that of the experiment. The simulated plate is $2\times2$~m$^2$ in physical units. Forcing is chosen to mimic the experimental one: the linear modes around $k=5\pi$ are forced resonantly at their linear frequency (close to 30~Hz as in the experiment). Linear dissipation is implemented in two ways: first a {\it Lorentz dissipation} (the dissipation time scale is decaying as a Lorentzian with the wavenumber $k$) as measured in the experiment~\cite{Miquel}. Such numerical simulations with realistic parameters have been shown to reproduce the observations for weakly non linear waves~\cite{Miquel2}.
Another dissipation (called {\it transparent dissipation} in the following) is implemented: dissipation is acting only at the highest wavenumbers of the simulation ($k>200\pi$). This enforces a constant flux of energy in the inertial (transparent) range between the forcing at $k\approx 5\pi$ and the dissipation at $k>200\pi$ consistent with the framework of the WTT. 

\paragraph{Analysis of Fourier spectra:} 

\begin{figure}[!htb]
\centering
\includegraphics[width=8.5cm]{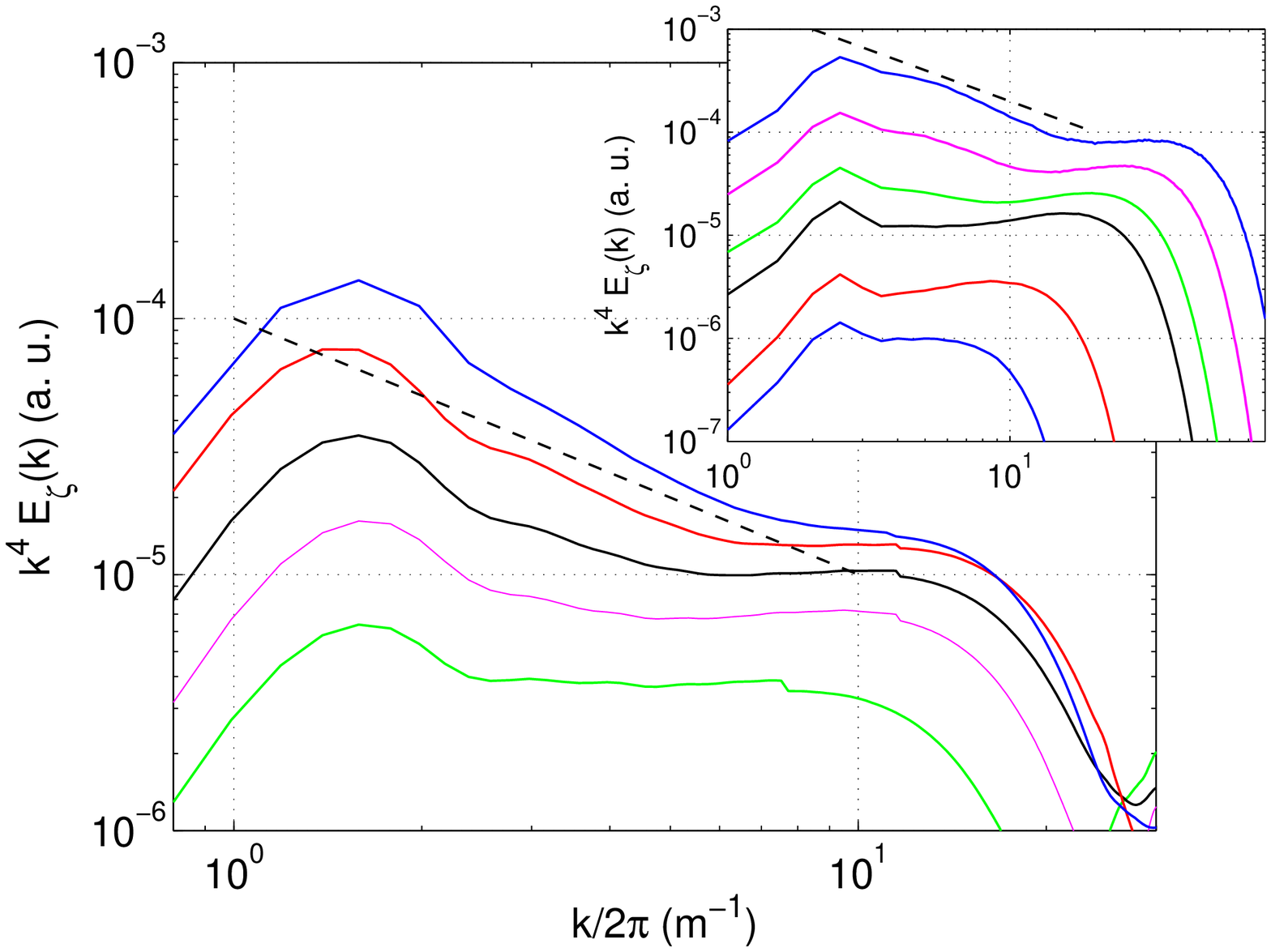}
\includegraphics[width=8.5cm]{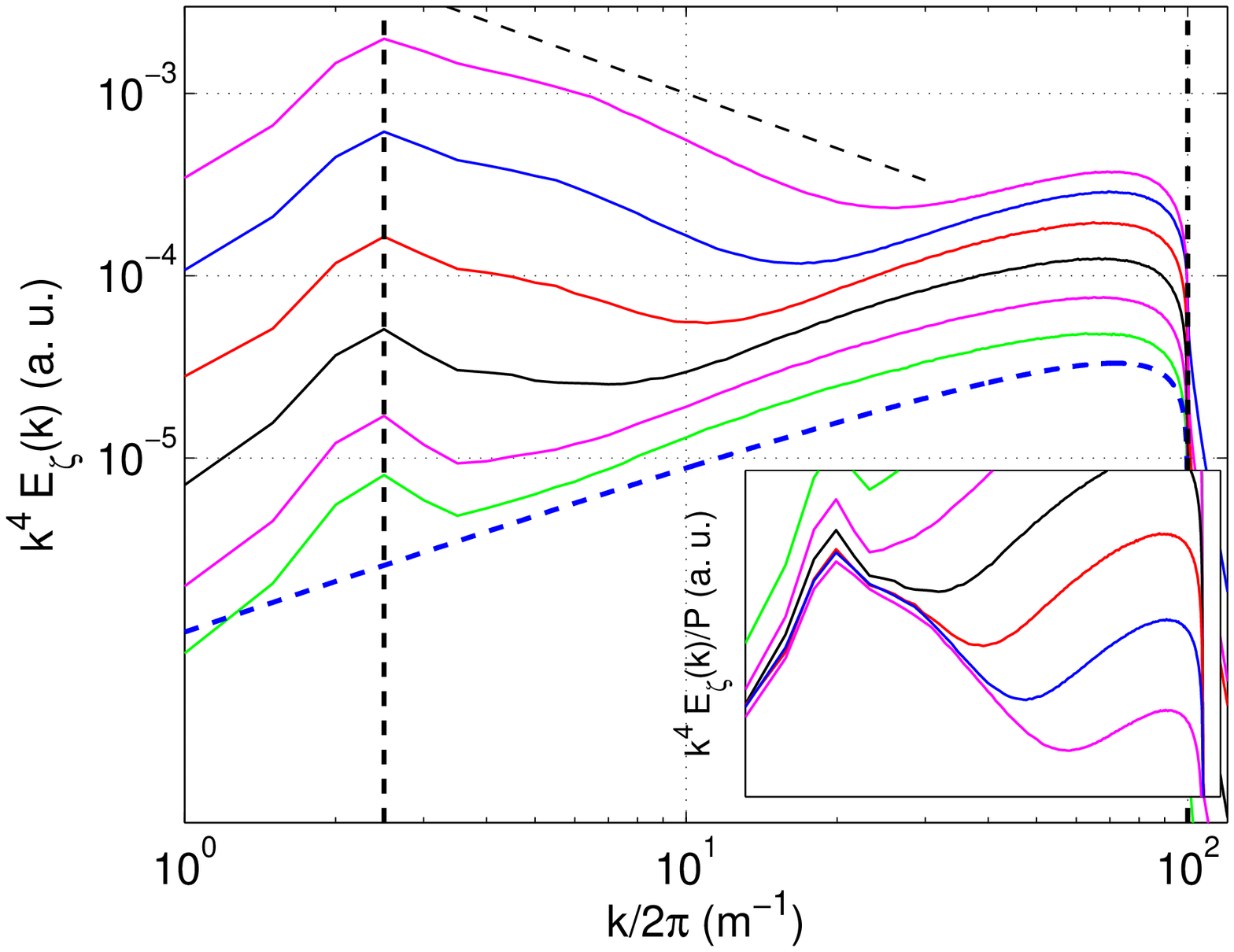}
\caption{Spectra of the plate deformation (multiplied by $k^4$). Forcing intensity is increasing from bottom curve to top curve. Top figure: experiment (forcing magnitude in arb. units: $0.5$, $0.75$, 1, $1.25$, $1.5$ from bottom to top). Inset: numerics with Lorentz dissipation mimicking the experimental dissipation (the forcing magnitude is increasing by a factor 2 between each curve from bottom to top). The dashed line is a $1/k$ decay. Bottom figure: numerical simulations with transparent dissipation. The left vertical line marks the forcing wavenumber and the right vertical line shows the lower wavenumber of the dissipation range. The lower dashed line stands for the theoretical prediction for weakly non linear wave turbulence~\cite{During}. The forcing magnitude is increasing by a factor 2 between each curve from bottom to top). Inset: spectrum rescaled by $P$.}
\label{sp}
\end{figure}
Figure~\ref{sp} shows power spectra of the deformation of the steel plate for several forcing intensities for experiment, numerical simulations with realistic dissipation and with transparent dissipation. All cases show the emergence of a new regime with a steeper spectrum at low wavenumber. The simulation shows a behavior very close to the experiment and is similar to the numerical results of~\cite{Yokoyama}. The experimental study confirms that the effect is present in real plates and is not an artifact of the F\"oppl-von K\'arm\'an equation driven out of their domain of validity. The simulation with transparent dissipation is in agreement with the theoretical prediction at low forcing intensity and shows the same departure at low wavenumber and strong forcing. Thus the observed change in spectrum is not due to a peculiar dissipation of real plates.

\paragraph*{Description of the new regime:} 
\begin{figure}[!htb]
\centering
\includegraphics[width=8.5cm]{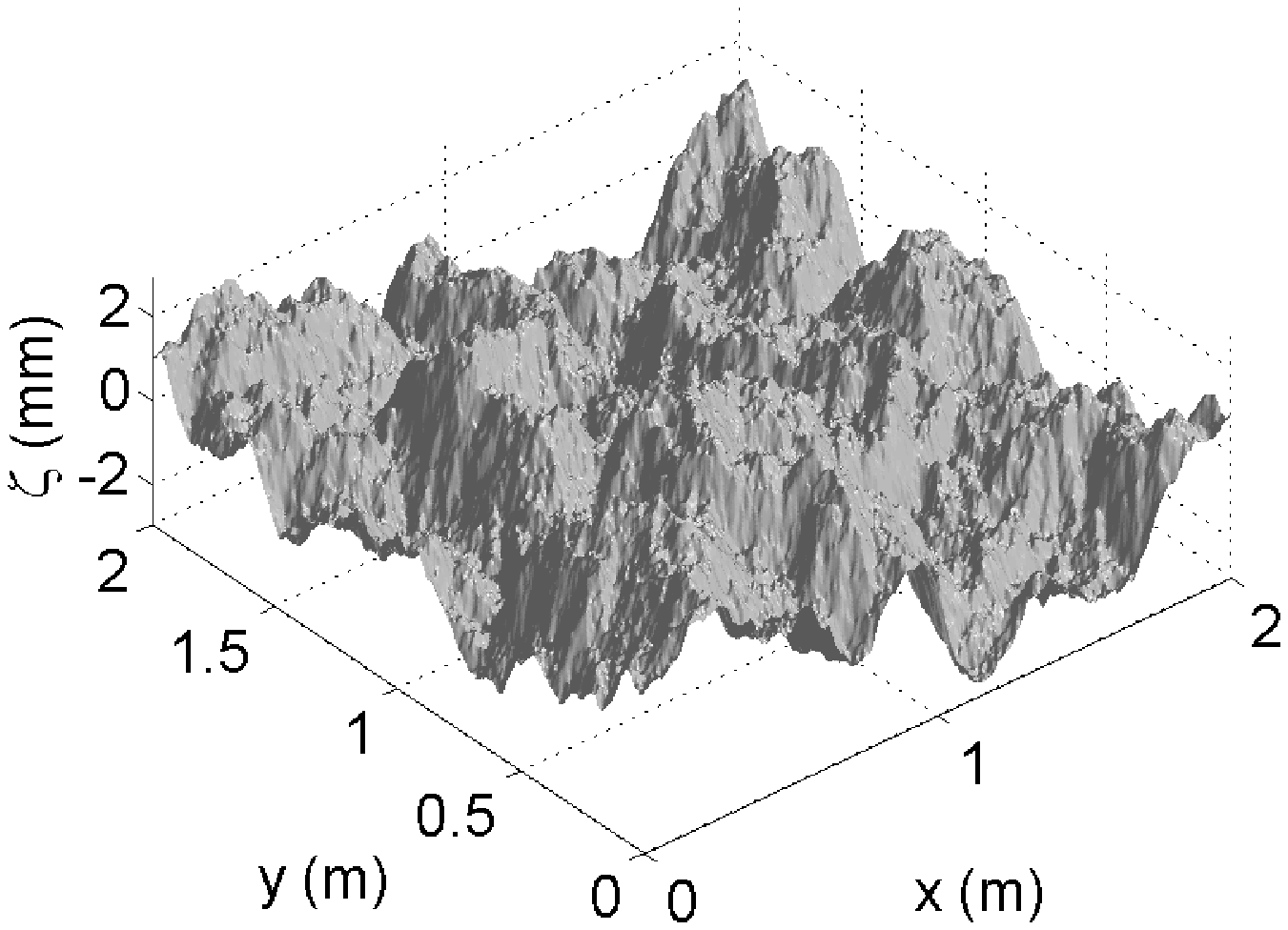}
\includegraphics[width=8.5cm]{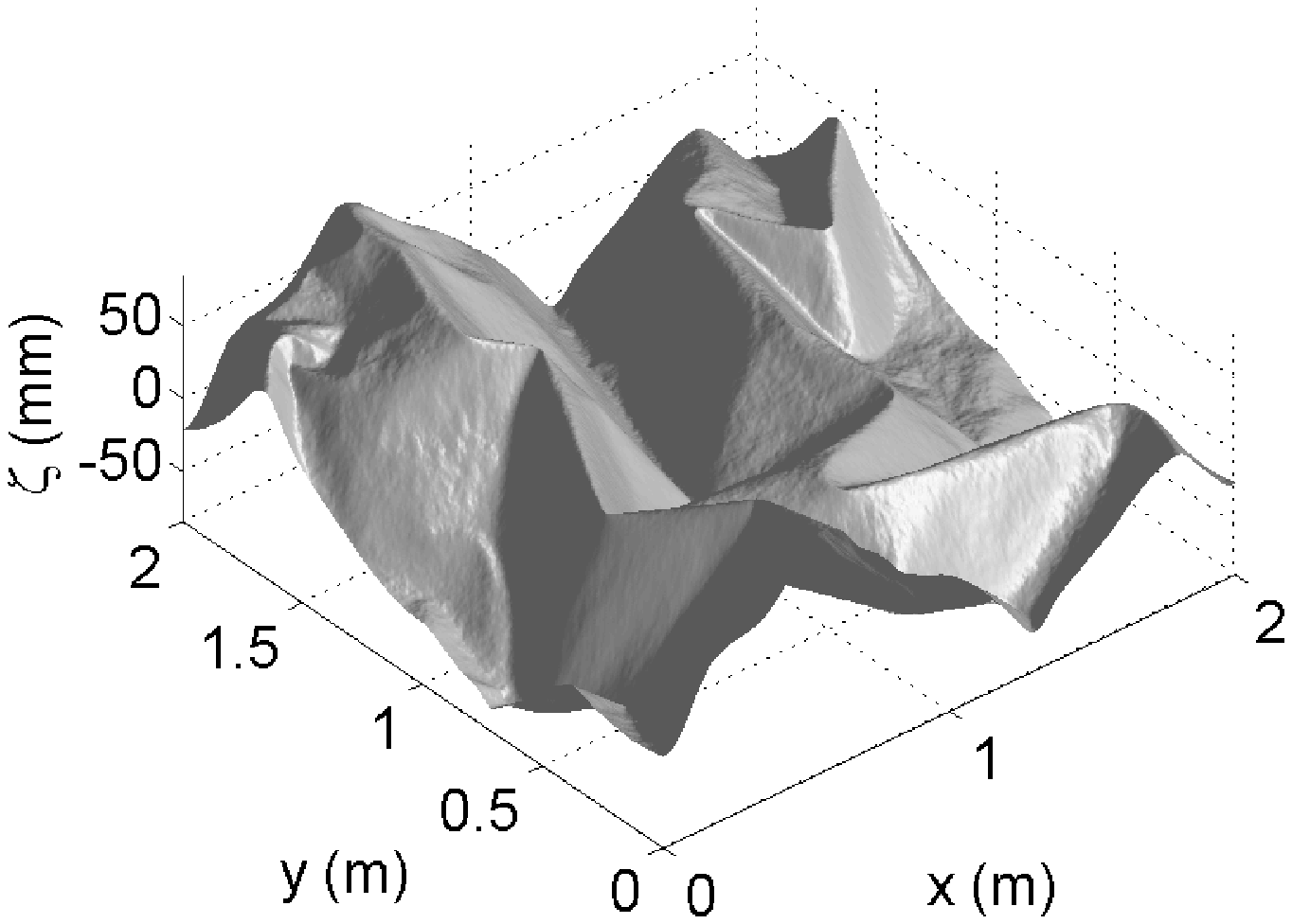}
\caption{Snapshots of the deformation for numerical simulations with transparent dissipation. Top: smallest forcing amplitude, bottom largest forcing amplitude.}
\label{def}
\end{figure}
Figure~\ref{def} shows snapshots of the deformation of the simulated plate in the case of transparent dissipation for the lowest and the strongest forcing 
intensities. The magnitude of the deformation is over one order of magnitude larger at the strongest forcing. The shape of the plate is very different in the two 
situations: at small forcing, where the KZ spectrum is observed, the deformation  looks as a rock surface and it appears extremely rough. On the other hand, 
at strong forcing, it resembles rather crumpled paper~\cite{RMPWitten} and seems smoother at first glance (note however the different vertical 
range between the two plots). The large scale deformation is made of folds with sharp crests that form ridges connecting developable cones 
(D-cones)~\cite{CerMa05}, responsible for the fast decay of the spectrum at small $k$. D-cones and ridges are singular structures that concentrate the stress resulting from the deformation of the plate. The difference between the two snapshots reveals a transition from a diffuse stress to a concentrated stress~\cite{RMPWitten} in a dynamical regime.
Indeed the folds are actually dynamical structures that forms, move 
for a while and then disappear: we refer to this new regime of vibration as {\it dynamical crumpling} by analogy 
with the {\it static} crumpling of an elastic sheet. The ridges and D-cones are regularized at small scale by the nonlinearity of 
the Von Karman equations~\cite{dcone}. In fact, as can be seen from the spectra in fig.~\ref{sp}, the small scale deformations are still following 
the KZ solution so that one can consider that small scale turbulent wave fluctuations still exist on top of the large scale folds. 

\begin{figure}[!htb]
\centering
\includegraphics[width=8.5cm]{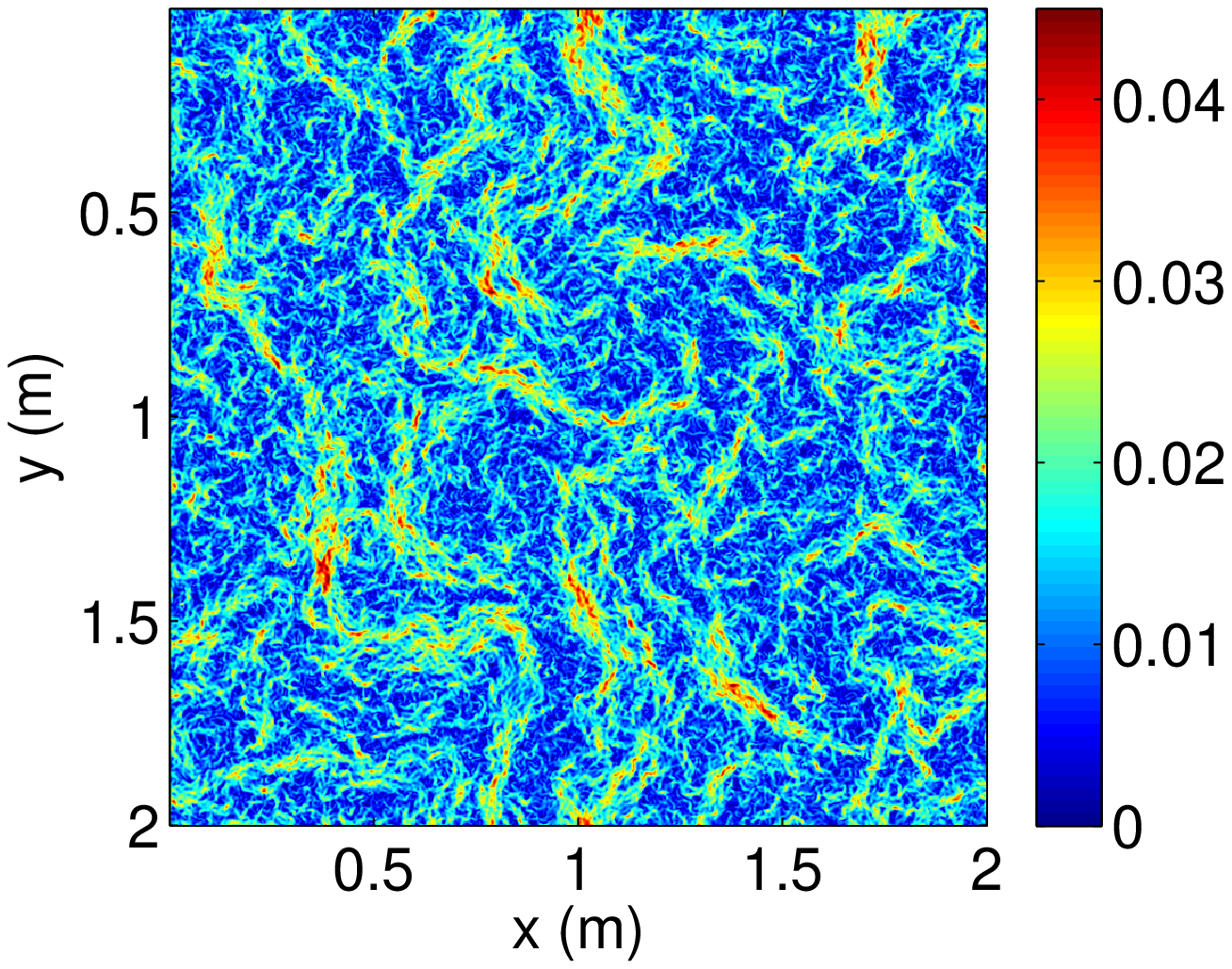}
\includegraphics[width=8.5cm]{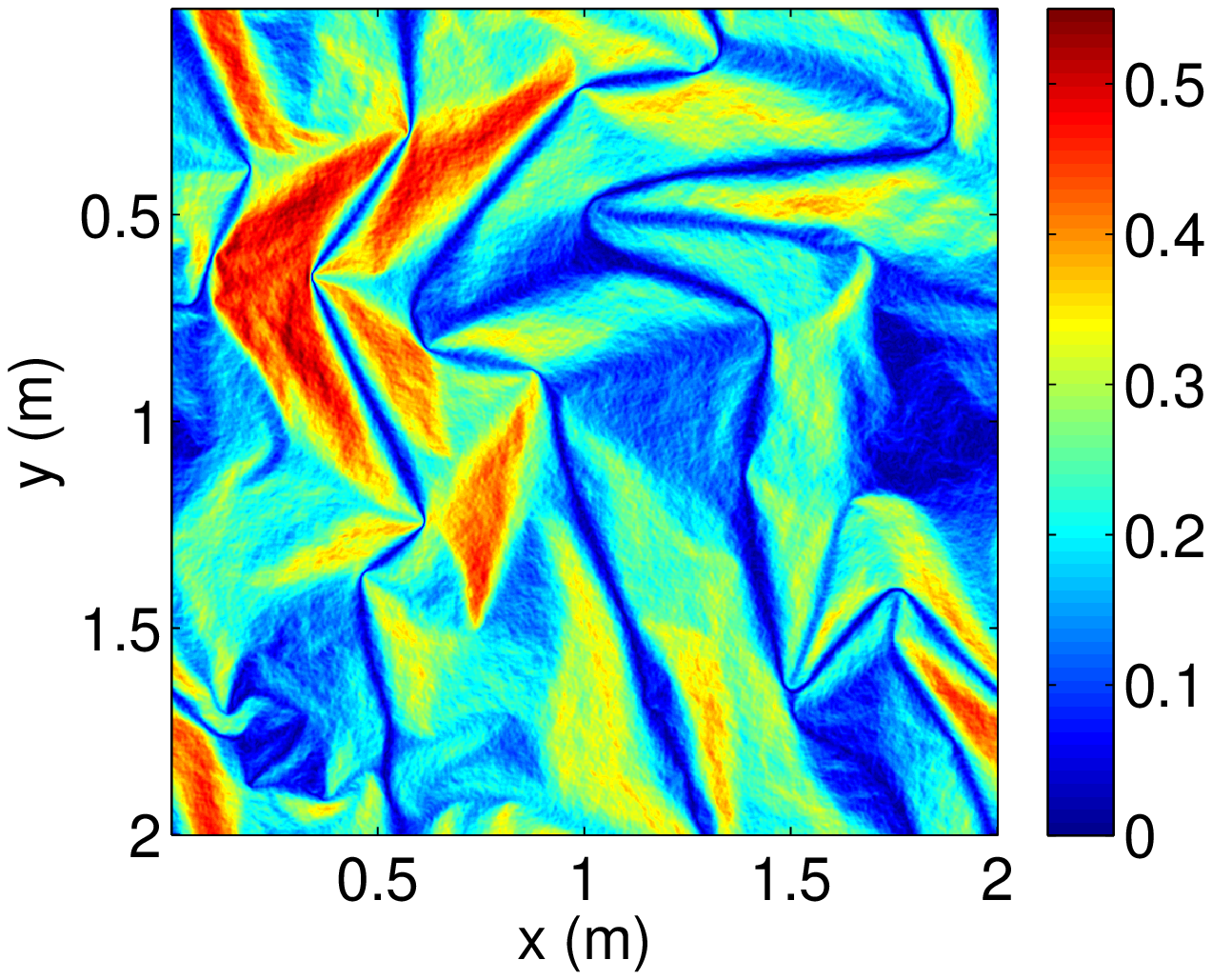}
\includegraphics[width=9cm]{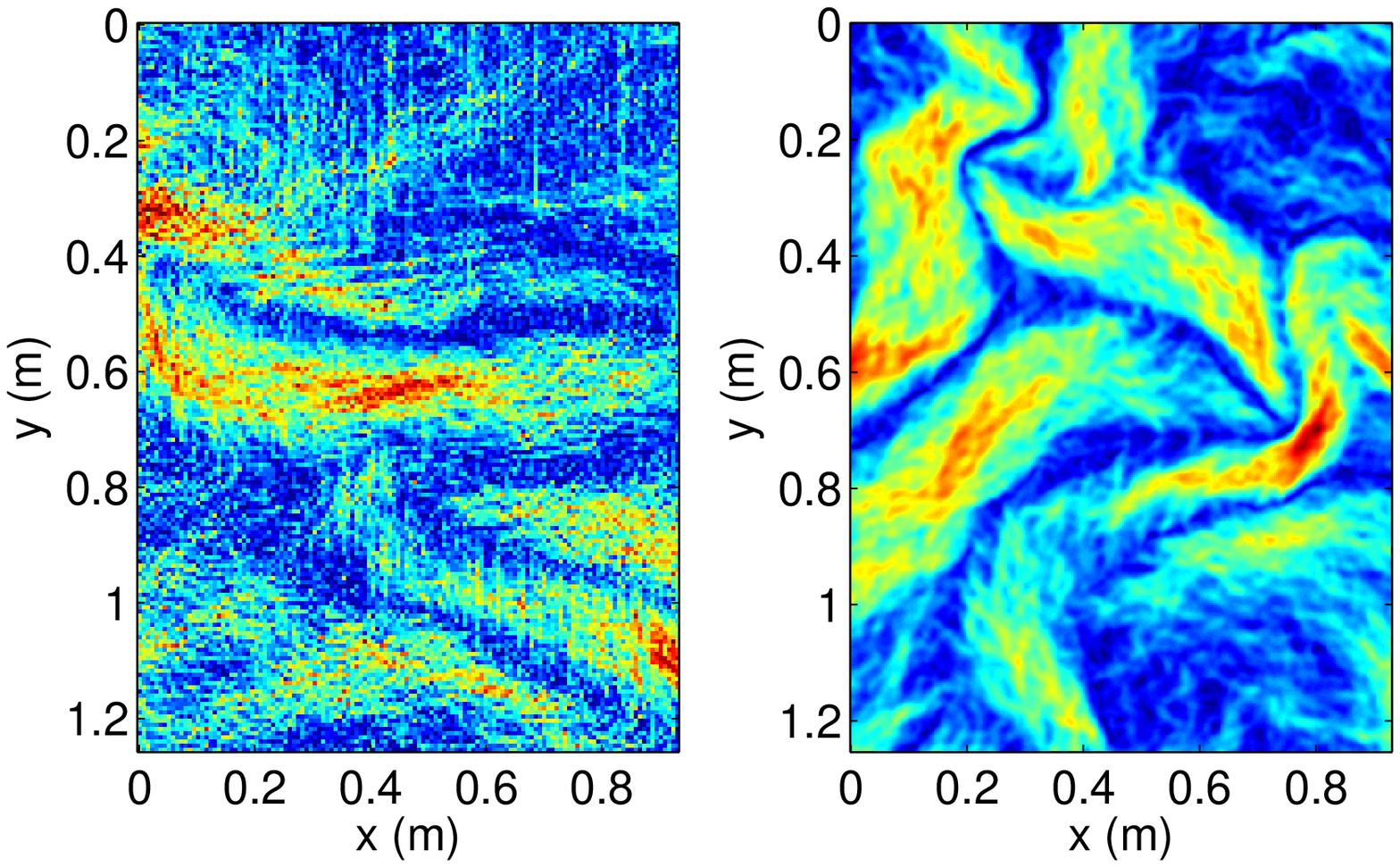}
\caption{Magnitude of the gradient of the deformation. Numerical simulations with transparent dissipation:  Top, smallest forcing amplitude. Center: largest forcing amplitude. Bottom: comparison of the experiment at strongest forcing (left) and numerical simulation with Lorentz dissipation (right). The simulation picture has been truncated to the size of the experimental picture.}
\label{grad}
\end{figure}
A more detailed observation of the deformation shows that on a single fold, the slopes are almost constant. This feature is also characteristic of static crumpling~\cite{RMPWitten}. This observation is supported by the pictures of the magnitude of the gradient displayed in fig.~\ref{grad} at the weakest and strongest forcing. At weak forcing, the picture of the gradient shows a small scale filamentary structure. At strong forcing, the gradient follows a very different spatial repartition: the largest values of the gradients are structured in stripes separated by very narrow ``valleys" of zero gradient corresponding to the ridges. 
The comparison between the experiment and the numerical simulations with the empirical Lorentz dissipation at strong forcing shows that similar features are observed for real plates.

\paragraph*{Analysis of spatial increments of the gradient}
\begin{figure}[!htb]
\centering
\includegraphics[width=8.5cm]{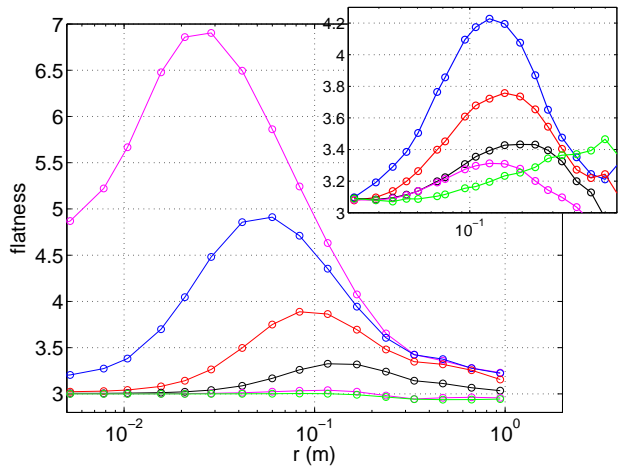}
\caption{Flatness of gradient increments (eq. (\ref{flate})). Main figure: numerics with transparent dissipation. Inset: experiment.}
\label{flat}
\end{figure}
A way to quantify the previous observations is to study the evolution of spatial increments of the spatial derivatives of the deformation:
\begin{equation}
\delta g_x(r)=\frac{\partial \zeta}{\partial x}(\mathbf R+r\hat{\mathbf x})-\frac{\partial \zeta}{\partial x}(\mathbf R)
\end{equation}
where $\hat{\mathbf x}$ is the unit vector along the $x$ direction and $\mathbf R$ a position in the $xy$ plane. We expect the probability distribution of such increments to change significantly between the weak forcing regime and the strong one reflecting the emergence of intermittency.
The flatness of the increments 
\begin{equation}
F(r)=\langle\delta g_x(r)^4\rangle/\langle\delta g_x(r)^2\rangle^2\, ,
\label{flate}
\end{equation} 
provides an indicator of the departure from a Gaussian statistics (for which the flatness is equal to $3$).
Figure~\ref{flat} shows the evolution of the flatness as a function of the forcing intensity for the experiment and for the numerics with transparent dissipation case (here the average is over $\mathbf R$ and time). At weak forcing, the flatness is close to 3 reflecting Gaussian statistics at all scales. Thus no intermittency is observed in this regime, the statistics are the same at all scales. At strong forcing, the flatness is close to 3 both at small and large scale but at intermediate scales of the order 10~cm the flatness takes values significantly larger than 3 (over 4 in experiment and over 7 in the numerics). The statistics are thus changing with scale reflecting the presence of intermittency. The length scale associated to the maximum flatness corresponds roughly to the width of the large gradient stripes observed in fig.~\ref{grad} (the forcing wavelength is about 40~cm). 
By contrast, the flatness of deformation or velocity increments (not shown) are almost unchanged when the forcing is increased and remain close to 3. 

\paragraph*{Discussion}
We have identified that the new dynamical regime is made of moving ridges delimited by D-cones.  It enters into the spectrum at small wave numbers 
(starting from the forcing scale) and invades more and more scales as the amplitude of the forcing increases. A surface dominated by ridges only 
would develop an elevation spectrum $E_\zeta(k)\propto 1/k^4$\cite{Kuz,During} {\it i.e.} slightly different from the KZ spectrum (\ref{denspec}) but not as steep as our observations. Modeling the D-cone by the simple deformation relation $\zeta=rf(\theta)$ (singular at 
$r=0$) yields the following contribution to the deformation spectrum~\cite{During}
 $E_\zeta(k)\propto 1/k^5,$
 in good agreement with the observed behavior at large scale as displayed in fig.~\ref{sp} where the slope $1/k$ is indicated for large forcing 
 amplitudes. The regularization of the singularity gives a correction to this scaling at short scale which is eventually not relevant in the observation
 since the spectrum is then dominated by the usual wave turbulence contribution. The observed intermittency is clearly related with the existence of these structures.
 
We would like to argue here that the emergence of this new dynamical regime is related to the breakdown of wave turbulence. A key ingredient in the WTT is that
the nonlinearities remain ``weak" at each scale or equivalently that the nonlinear time scale is much larger than the linear one. Since the magnitude of 
the wave amplitude evolves with the scale within the cascade, the ratio between these two timescales usually vary between the forcing and the 
dissipative scales. Then, if the inertial range is wide enough the non linearity becomes significantly strong and intermittency develops either at the 
largest or smallest scales~\cite{ZakNew,NNB}. This regime of strong turbulence due to the breakdown of weak turbulence leads to different dynamical 
properties for which a general theory is still missing. Comparing the
linear time ($\tau_L$ due to the wave oscillation) to the nonlinear time ($\tau_{NL}$ due to the nonlinear interactions between the waves) at each 
scale for the KZ spectrum yields, following~\cite{NNB,During} (up to the logarithmic correction terms)
\begin{equation}
 \frac{\tau_L}{\tau_{NL}} \sim \frac{ P^{2/3}}{k^4}.
 \end{equation}
The wavenumber scaling of this ratio shows that the wave turbulence hypothesis of time separation between the fast wave oscillation and asymptotic nonlinear contribution
fails at large scale and for high forcing amplitude. Here the singularities dominate the spectrum at low wave number consistently with the above argument. At higher wave numbers the scale separation $\tau_{NL}\gg\tau_L$ is still satisfied so that weak turbulence can proceed. 
The spectrum of the new regime is scaling as $P/k^5$ as seen in fig.~\ref{sp} (bottom) suggesting that the number of singularities increases also linearly with $P$~\cite{Kuz}.

Finally it can be noted that experiments on water surface waves (gravity-capillary waves) in which singularities are indeed observed~\cite{Herbert} also show a spectrum that scales linearly with $P$ in place of the Kolmogorov-Zakharov scaling~\cite{Falcon,Berhanu} ($P^{1/3}$ for gravity, $P^{1/2}$ for capillary). It may be due to the fact that the observed regime of water waves is dominated by singularities, consistent with the observation that the scaling in wavenumber is not the one predicted by WTT for water waves.  

\begin{acknowledgments}
This research was supported by the ANR grant TURBULON 12-BS04-0005.
\end{acknowledgments}

\bibliography{biblio7}

 \end{document}